\documentclass[aps,prl,floatfix,twocolumn,reprint,amsmath,amssymb,superscriptaddress]{revtex4-1}
\usepackage{amsfonts}
\usepackage{mathrsfs}
\usepackage{amsmath}% needed for subequations
\usepackage{color}
\usepackage{natbib}
\usepackage{graphicx}
\usepackage{bm}% bold maths
\usepackage{amssymb}
\usepackage{xspace}
\usepackage{epstopdf}
\usepackage{dcolumn}% Align table columns on decimal point
\usepackage{longtable}
\usepackage{multirow}
\usepackage[colorlinks=true, letterpaper=true, pdfstartview=FitV, linkcolor=blue, citecolor=blue, urlcolor=blue]{hyperref}
\UseRawInputEncoding

\makeatletter

\newcommand{\Rmnum}[1]{\expandafter\@slowromancap\romannumeral #1@}
\makeatother

\begin{document}

\title{Theoretical Design of Mono-Elemental Ferroelectricity with Tunable Spin Textures in Bilayer Tellurium}
\author{Jiajun Zhu}

\affiliation{Yunnan Key Laboratory of Electromagnetic Materials and Devices, National Center for International Research on Photoelectric and Energy Materials, School of Materials and Energy, Yunnan University, Kunming, 650091,P.R.China}
\author{Botao Fu}
\email[]{fubotao2008@gmail.com}
\affiliation{College of Physics and Electronic Engineering, Center for Computational Sciences, Sichuan Normal University, Chengdu, 610068,
P.R.China}
\author{Heyun Zhao}
\affiliation{Yunnan Key Laboratory of Electromagnetic Materials and Devices, National Center for International Research on Photoelectric and Energy Materials, School of Materials and Energy, Yunnan University, Kunming, 650091,P.R.China}
\author{Wanbiao Hu}
\email[]{huwanbiao@ynu.edu.cn}
\affiliation{Yunnan Key Laboratory of Electromagnetic Materials and Devices, National Center for International Research on Photoelectric and Energy Materials, School of Materials and Energy, Yunnan University, Kunming, 650091,P.R.China}
\affiliation{Electron Microscopy Center, Yunnan University, Kunming 650091, P.R.China}

\date{\today}

\begin{abstract}
2D Ferroelectricity with switchable electric polarization has drawn widespread attention in condensed matter physics due to its crucial applications in non-volatile memory and ferroelectric spin devices. Despite recent progress in 2D ferroelectric, achieving the mono-elemental ferroelectricity still remains a great challenge because most nonmetallic mono-elemental materials are stabilized in nonpolar crystal structures.
In this work, we theoretically designed mono-elemental ferroelectricity with tunable and significant spin textures in bilayer tellurium (BL-Te). Comprehensive quantitative polarization calculations demonstrate that asymmetric stacking in BL-Te can generate out-of-plane (OOP) polarization with a magnitude of 0.78 pC/m. This polarization stems from distinguishing interlayer and intra-layer contributions.
Moreover, these stacked BL-Te, characterized by significant spin-orbit coupling, serve as an ideal platform for investigating both conventional spin polarization and layer-dependent/hidden spin polarization through ferroelectric reversion. Our work not only broaden the category of 2D mono-elemental ferroelectric but also offer a new platform for multifunctional nanodevices.

\end{abstract}

\maketitle

\begin{figure*}
	\includegraphics[width=17cm]{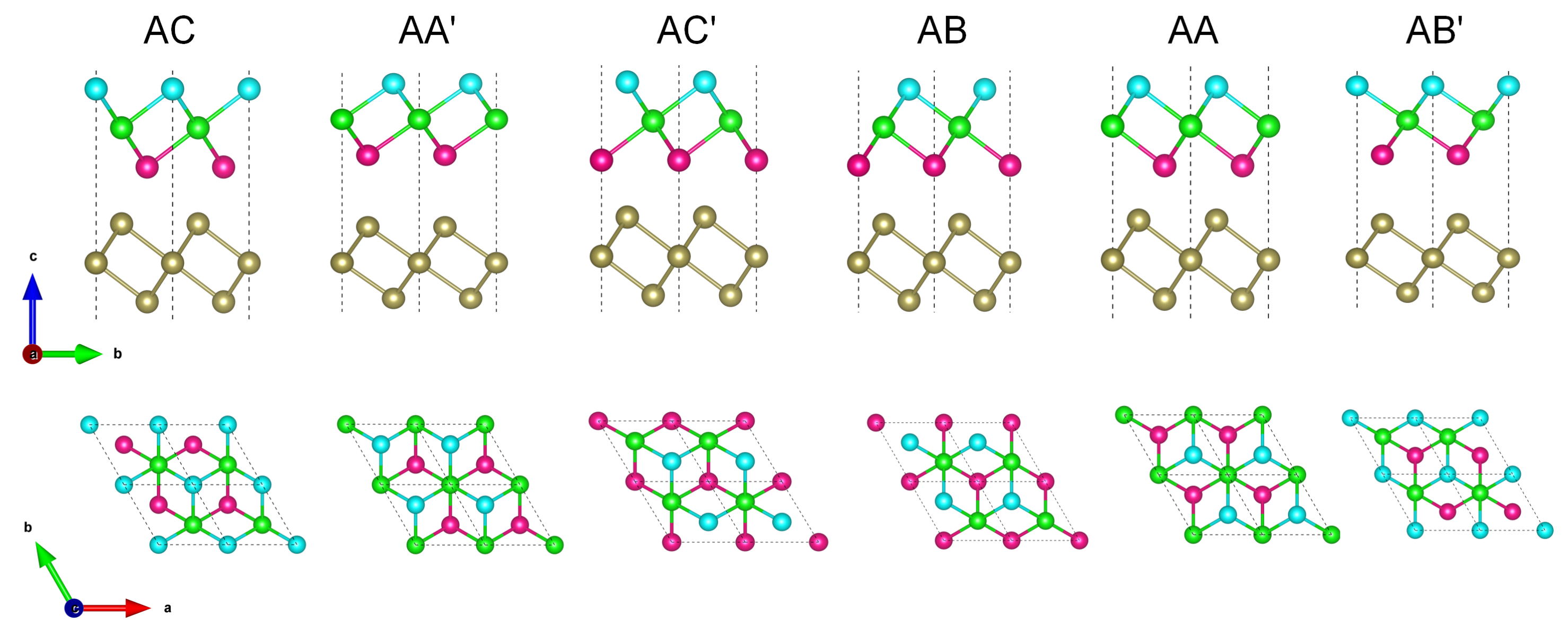}
	\centering
    \caption{  Side and top views of six high-symmetry stacking configurations of BL-Te (AC, AA$^{\prime}$, AC$^{\prime}$, AB, AA, AB$^{\prime}$).
    }
    \label{fig1}
\end{figure*}

Recently, two-dimensional (2D) ferroelectric materials have garnered significant attention due to their immunity to size-related effects and their ability to exhibit electric polarization at the monolayer level\cite{junquera2003critical,fong2004ferroelectricity}.
At first, such 2D ferroelectricity was theoretically proposed\cite{PhysRevLett.130.146801,PhysRevLett.131.096801} and experimentally verified among very limited 2D materials  featured with spontaneously centro-symmetric breaking and polar point group, such as $\alpha$-In$_{2}$Se$_{3}$\cite{cui2018intercorrelated,xiao2018intrinsic}, CuInP$_{2}$S$_{6}$\cite{liu2016room}, d1T-MoTe$_{2}$\cite{yuan2019room}, $\alpha$-phase group IV-VI monochalcogenides\cite{PhysRevLett.117.246802,wu2016intrinsic,PhysRevB.97.024110}.
Soon after, sliding ferroelectricity was proposed by Wu et al\cite{li2017binary}, which can be widely achieved through non-symmetric stacking of two otherwise non-polar monolayers\cite{yuan2023probing}. This sliding ferroelectricity was experimentally observed in $h$-BN\cite{yasuda2021stacking,vizner2021interfacial,tsymbal2021two} and MoS$_2$ series \cite{deb2022cumulative,yang2018origin,jindal2023coupled}.
Bearing that the sliding ferroelectricity vdW bilayer has a lower ferroelectric transition barrier and various choice of potential materials due to powerful heterojunction technology\cite{PhysRevApplied.19.034002,PhysRevB.107.035426,PhysRevLett.128.067601}, it will offers greater potential for multiferroic coupling compared to conventional 2D ferroelectric materials\cite{wu2023coexisting,PhysRevB.107.195128,marmolejo2022slippery}.

On the other hand, 2D mono-elemental materials (termed as Xenes), have consistently garnered significant attention due to their simple composition and diverse range of physical properties. However, achieving ferroelectricity in mono-elemental materials is inherently challenging because most nonmetallic mono-elemental materials are stabilized in nonpolar crystal structures of high symmetry. Nevertheless, two mechanisms have emerged to modifying the equivalence or symmetry of the same elements, which can be further employed to realize mono-elemental ferroelectricity: (1) In mono-elemental monolayer containing unstable lone-paired electrons, intrinsic structural distortion occurs\cite{xiao2018elemental,wang2018two,doi:10.1021/acs.nanolett.2c04723}, which may lead to spontaneous ferroelectric polarization, as experimentally confirmed in monolayer $\alpha$-Bi\cite{gou2023two}; (2) In van der Waals bilayer Xenes, switchable electric polarization is achieved through inter-layer sliding between two non-polar parent monolayers\cite{liang2021out,wang2023towards}, known as sliding mono-elemental ferroelectricity. It's worthwhile that very recently such sliding mono-elemental ferroelectricity was successfully observed in multiple layer graphene, offers a promising platform for advancing the field of mono-elemental ferroelectric.
Nevertheless, we notice that (i) presently mono-elemental ferroelectric materials remain scarce\cite{PhysRevApplied.11.064063}, especially those with substantial spin-orbit coupling effects; (ii) intriguing interaction between ferroelectric polarization, spin polarization, layer-dependent physics still remains unexplored.

Motivated by these background,  we theoretically predict the emergence of sliding ferroelectricity in newly a synthesized 2D tellurene and demonstrate the tunable spin textures by flexibly switching the electric polarization. Beginning with six distinct bilayer stacking configurations (AC/AC$^{\prime}$, AB/AB$^{\prime}$, AA/AA$^{\prime}$), structural evolution between them is clearly illustrated through interlayer sliding and rotation. Subsequently, we confirm that the ground state (AC/AC$^{\prime}$ phases) hosts a polar point group that enables the occurrence of ferroelectricity. Employing Berry phase approach and modern ferroelectric theory\cite{PhysRevB.47.1651}, an electric polarization of 0.78 pC/m is obtained in BL-Te, which is generally attributed to inter- and intra-layer contributions originated from the interlayer van der Waals forces that cause local interface asymmetry and global polar point group.
Moreover, with substantial spin-orbital coupling effect, the BL-Te demonstrates switchable spin polarizations for opposite FE phase and layer-dependent spin texture for paraelectric phase.  Our work contributes to fundamental research in ferroelectrics and spintronics.

\begin{figure*}[htp]
	\includegraphics[width=17cm]{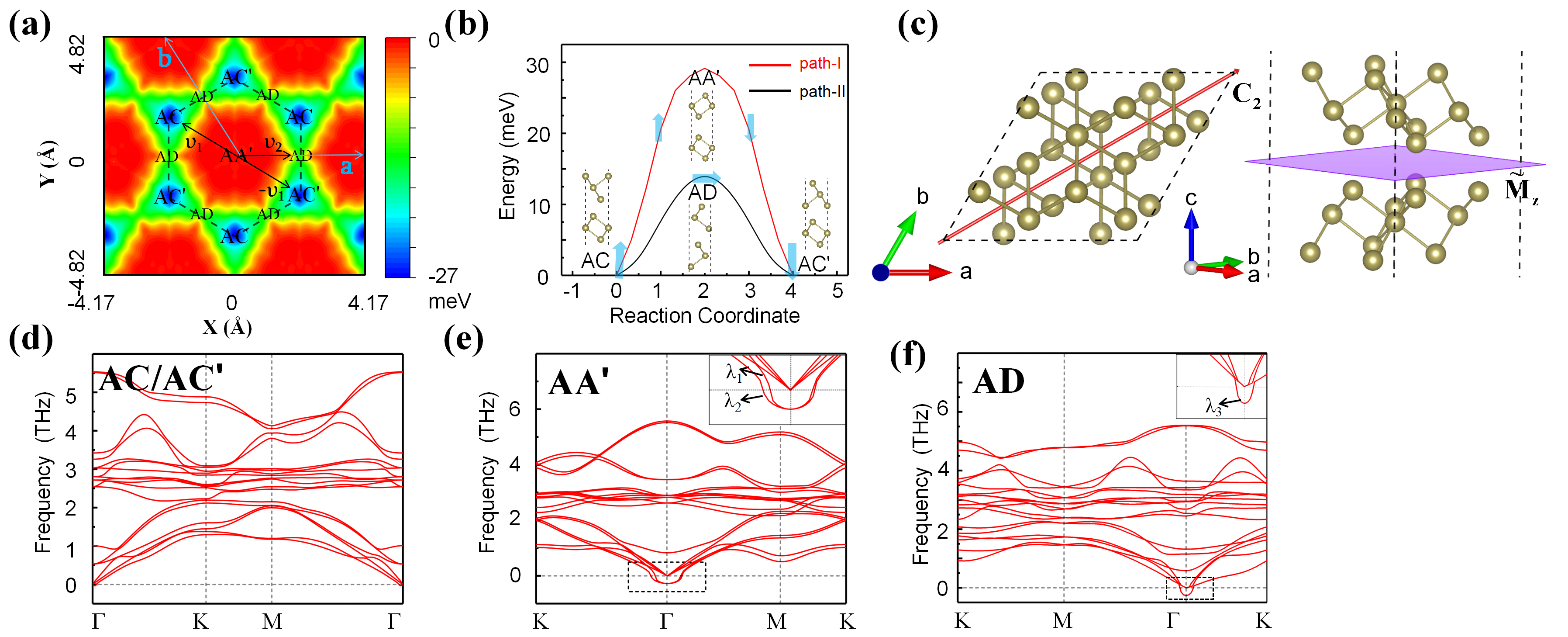}
    \caption{ (a) Energy barrier diagram for AA$^{\prime}$, AC and AC$^{\prime}$ stacking of BL-Te, the AC, AC$^{\prime}$ and AA$^{\prime}$ can be sliding each other£¬ where \boldsymbol{$\nu_1$}=(-$\frac{1}{3}$\textbf{a},$\frac{1}{3}$\textbf{b}) and \boldsymbol{$\nu_2$}=($\frac{1}{2}$\textbf{a},0). (b) Energy pathway of ferroelectric switching as a function of step number within the CI-NEB. (c) the side and top view for AD phase, the blue xy plane represents a slide mirror $\widetilde{M}$$_{z}$. (d-f) Phonon spectra for AC and AA$^{\prime}$ and AD phase. The symbol $\lambda$$_1$, $\lambda$$_2$ and $\lambda$$_3$ indicate the soft optical modes.
    }
    \label{fig2}
\end{figure*}
%%%%%%%%%%%
 Recently, a new category of monolayer tellurene (ML-Te) was successfully synthesized on pyrolytic graphite substrates\cite{PhysRevLett.119.106101}. This material exhibits three distinct crystal structures: the most stable 1T-MoS$_2$-like ($\alpha$-Te), a metastable tetragonal structure ($\beta$-Te), and a 2H-MoS$_2$-like($\gamma$-Te). In this study, we specifically focus on the ground state $\alpha$-Te, which is a semiconductor with a band gap of 0.460 eV, similar to that of its bulk counterpart. Unless stated otherwise, when we refer to ``tellurene", we are referring to $\alpha$-Te. ML-Te crystals in a hexagonal lattice with three sub-layers located at A, B, and C sites, respectively. Notably, this structure is centrosymmetric with respect to the central Te atom, forbidding the emergence of ferroelectricity. Fortunately, stacking engineering has emerged as an effective and powerful technique for manipulating the structural and electronic properties of various 2D van der Waals (vdW) materials\cite{zhang2023ferroelectric}. Different stacking methods can yield diverse space groups. Thus, by employing specific stacking patterns, most 2D monolayers can form corresponding bilayers with noncentrosymmetric space groups and polar points. This provides a promising strategy for designing new ferroelectric materials.

In the case of BL-Te, within its hexagonal lattice, we explored all high-symmetry stacking possibilities, resulting in the creation of six distinct bilayer configurations, as illustrated in Fig.~\ref{fig1}. These BL-Te can be divided into two groups based on the presence or absence of spatial inversion symmetry. The first group is non-centrosymmetric, comprising the AC/AC$^{\prime}$ phases hosting the space group $P3m1$ (No.156) and polar point group $C_{3v}$ and the AA$^{\prime}$ phase with space group $P\overline{6}m2$ (No.187) and nonpolar point group $D_{3h}$. The second group is centrosymmetric including AB/AB$^{\prime}$ and AA phases that share the same space group $P\overline{3}m1$ (No.164) with nonpolar point group $D_{3d}$. More interesting, the materials in each group can evolve with each other through in-plane slide, and the between two group can evolve with each other through vertical rotation [see Fig. S1]. Given the symmetry constraints associated with ferroelectric materials, our primary focus centers on the three non-centrosymmetric structures.

Commencing from the AA$^{\prime}$ stacking structure characterized by M$_z$ symmetry forbidding out-of-plane (OOP) ferroelectricity, we initiate a sliding transformation of the upper layer along the \textbf{a} and \textbf{b} axes using the sliding vector \boldsymbol{$\nu$}=(\emph{l}$_a$\textbf{a}, \emph{l}$_b$\textbf{b}), where \textbf{a} and \textbf{b} represent the basis vectors of the crystal. The resulting energy contour is depicted in Fig.~\ref{fig2}(a). Evidently, the central point, representing the AA$^{\prime}$ phase, corresponds to the energy maximum.
Surrounding the central point, six energy minima are symmetrically located at the vertex of a hexagon, identified as AC phase with a slide vector \boldsymbol{$\nu_1$}=(-$\frac{1}{3}$\textbf{a},$\frac{1}{3}$\textbf{b}) and AC$^{\prime}$ phase with a slide vector -\boldsymbol{$\nu_1$}=($\frac{1}{3}$\textbf{a},-$\frac{1}{3}$\textbf{b}). Obviously, they are spatial inversion of each other. Moreover, the AC/AC$^{\prime}$ phase possesses the same polar point group, C$_{3v}$, which breaks M$_{z}$ symmetry, allowing the appearance of OOP ferroelectricity. Therefore, we predict that the AA$^{\prime}$ phase at the apex of potential energy surface is energy unstable, which will spontaneously transform into the energy favorable AC/AC$^{\prime}$ phase, and meanwhile accompanied by the M$_z$ symmetry breaking and the emergence of OOP ferroelectricity. As demonstrated in Fig.~\ref{fig2}(b), the transition barrier from the ferroelectric AC/AC$^{\prime}$ phase to the paraelectric AA$^{\prime}$ phase is calculated by CI-NEB\cite{10.1063/1.1329672}, which is about 29.2 meV. This path is termed as ``path-I".

Moreover,as shown in Fig.~\ref{fig2}(a), we found that there are six saddle points at the midpoint on the hexagonal edge, denoted as AD phase.
This AD phase can connect the neighboring AC and AC$^{\prime}$ phase through a sliding vector \boldsymbol{$\nu$$_2$}= ($\frac{1}{2}$\textbf{a},0). The Fig.~\ref{fig2}(c) illustrates that the AD phase exhibits a lower symmetry, featuring a nonsymmorphic space group of $Abm2$(No.39).
It has a slide mirror $\widetilde{M}$$_{z}$ plane that forbid the OOP FE while has the polar point group of $C_{2v}$ that allows for the in-plane  electric polarization along the diagonal direction. In Fig.~\ref{fig2}(b), the transition path from ferroelectric AC/AC$^{\prime}$ phase to AD phase is calculated, which give a transition barrier of only 13.9 meV. This path is termed as ``path-II".
Remarkably, the transition path-II offers a shorter spatial distance and a smaller energy barrier compared to path-I, suggesting that it represents the most favorable ferroelectric flipping pathway. Similar transition path is also discovered proposed in bilayer MnBi$_2$Te$_4$\cite{PhysRevB.108.104109}. In particular, it's worthwhile the middle state AD phase itself has in-plane electric polarization (9.8 pC/m) along transition direction, which is beneficial for the electric filed induced inter-layer sliding and flipping of OOP FE.

The above FE transformation process are also clearly demonstrated through the analysis of the phonon spectra as illustrated in Fig.~\ref{fig2}(d)-(f). It is evident that the AC/AC$^{\prime}$ phase exhibits no imaginary frequencies across the entire Brillouin zone (BZ) , confirming its dynamically stability in Fig.~\ref{fig2}(d). In contrast, the AA$^{\prime}$ exhibit two soft optical phonon modes ($\lambda$$_1$ and $\lambda$$_2$) around $\Gamma$ point in Fig.~\ref{fig2}(e). Further vibrational vector analysis in Fig. S2(e) reveals that the soft mode $\lambda$$_1$ corresponds to interlayer sliding in the \boldsymbol{$v_1$} direction, while the soft mode $\lambda$$_2$ corresponds to interlayer sliding in the -\boldsymbol{$v_1$} direction. Namely, AA$^{\prime}$ phase is dynamically unstable and will spontaneous transform into dynamically stable AC/AC$^{\prime}$ phase in consist with preceding conclusion. Similar analysis also works for AD phase, revealing the structural transition form AD into AC/AC$^{\prime}$ phase in Fig.~\ref{fig2}(f).

\begin{figure}
	\includegraphics[width=8.5cm]{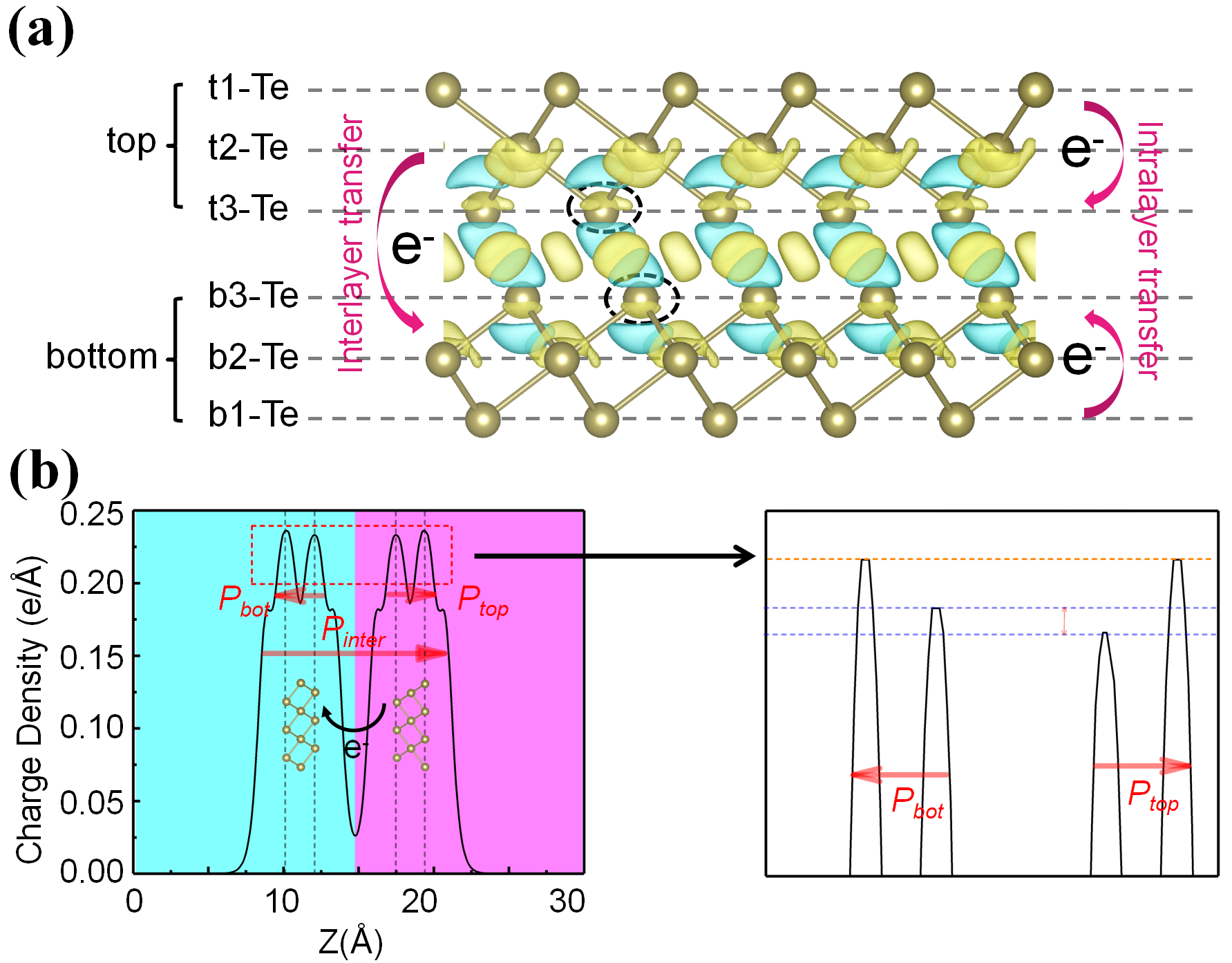}
    \caption{ (a) Charge density difference for AC phase of BL-Te, where the yellow and blue areas represent electron accumulation and depletion. (b) The planar averaged differential charge density ($\Delta$$\rho$) of AC phase along the z direction. The blue and purple areas show the charge density of the bottom and top layers, the red arrow indicates electrical polarization.
     }
    \label{fig3}
\end{figure}

%%%%%%%%%%%%%%%%
\begin{table}
  \centering
  \centering
\caption{Calculated electric polarization ($P_s$) for mono-elemental ferroelectric materials}
\setlength{\tabcolsep}{0.05cm}{
\begin{tabular}{cccc}
\hline
Materials   & $P_s$ (pC/m) & Materials & $P_s$ (pC/m)   \\
\hline
ML $\alpha$-As\cite{xiao2018elemental}  & 0.46    & ML $\alpha$-Sb\cite{xiao2018elemental}          & 0.75                 \\
ML $\alpha$-Bi\cite{xiao2018elemental}  & 1.51    & BL $\beta$-P\cite{liang2021out}        & 0.53                \\
BL $\beta$-As\cite{liang2021out}& 1.33   &  BL $\beta$-Sb\cite{liang2021out}       & 1.45                 \\
BL $\alpha$-Te        & 0.78             &  BL $\beta$-Te\cite{wang2018two}     & 1.02                \\
Few-layer graphene \cite{PhysRevLett.131.096801}      & 0.17-0.32  \\
\hline
\end{tabular}} \label{tab1}
\end{table}

The electric polarization intensity $P_s$ in the periodic system was evaluated by using the Berry phase method\cite{PhysRevB.47.1651}. The calculated $P_s$ for BL-Te with AC/AC$^{\prime}$ phase is about $\pm$0.78 pC/m along $z$-direction, which is comparable with that in other reported mono-elemental materials as listed in Tab.~\ref{tab1}.
To explore the mechanism of polarization in BL-Te, the differential charge density ($\Delta$$\rho$) of BL-Te respect to isolated top and bottom tellurene monolayers is calculated and shown in Fig.~\ref{fig3}(a). It's knowm that for an isolated ML-Te, the geometric structure and charge distribution of t1 layer and t3 layer are symmetric with respect to t2 layer, indicating no net residual polarization. However, in BL-Te, the electrons evidently transfer from t3 to t1 in top layer and transfer from b3 to t1 in the bottom layer, disrupting the initial symmetry within in solely monolayer. This asymmetry arises due to the differing environments faced by t1/t3 and b1/b3 layers in the bilayer, disrupting their natural center symmetry compared to the isolated monolayer.
Consequently, due to interface effects the intra-layer charge transfer induces asymmetry in the electron distribution, leading to intra-layer polarization, denoted as $P_{bot}$ and $P_{top}$ for bottom and top tellurenes respectively.
Furthermore, due to the asymmetrical interlayer stacking (breaking M$_{z}$) for AC/AC$^{\prime}$ phase,  it is evident that the charge distribution between the top and bottom layers is asymmetric. Thus, there must be charge transferring between the top and bottom layers, as illustrated in the Fig.~\ref{fig3}(b), where electrons move from the bottom layer to the top layer. This further leads to the interlayer electric polarization, denoted as $P_{inter}$.

Based on above analysis, the overall polarization in BL-Te, $P_s$ can be decomposed into interlayer polarization ($P_{inter}$) and intralayer polarization contributions ($P_{bot}/P_{top}$).
Based on the definition of polarization for a finite system\cite{spaldin2012beginner},
\begin{align}
{\mathbf{P}}=\int \mathbf{r} \rho (\mathbf{r}) \mathrm{d}\mathbf{r},
\end{align}
the total polarization strength ($P_s$) and its different components can be quantitatively calculated.
In the plotted electron distribution function $\rho (z)$ along the z-direction, it is evident that the electron distribution between the bottom and top layers is asymmetric. Using the integral in the equation (1), we obtain polarizations for the top and bottom layers as $P_{top}=1.0$ pC/m and $P_{bot}=-0.8$ pC/m, respectively.
Additionally, integrating the charge density for each layer allows us to estimate the interlayer charge transfer $\delta q$ between the two layers to be approximately 0.005 $e$. Consequently, the interlayer polarization induced by this charge transfer is estimated to be $P_{inter}=0.6$ pC/m.

Quantitatively comparing the overall polarization strength of BL-Te with the specific contributions from its three parts, we observe that: (1)
The contributions of these three part are all significant, playing crucial roles to the overall polarization at a comparable magnitude; (2) The polarizations induced by the top and bottom layer energies have opposite directions; (3) The interlayer contribution determines the overall direction of the system's polarization.
The total polarization magnitude ($P_s$) can be considered as jointly determined by these three components.
Besides, for AD and AA$^{\prime}$ phases with mirror symmetry, the $\delta q$ is forced to be zero and $P_{bot}$=-$P_{top}$, which leads to $P_s=0$ as expected.

\begin{figure*}[ht]
   \includegraphics[width=18cm]{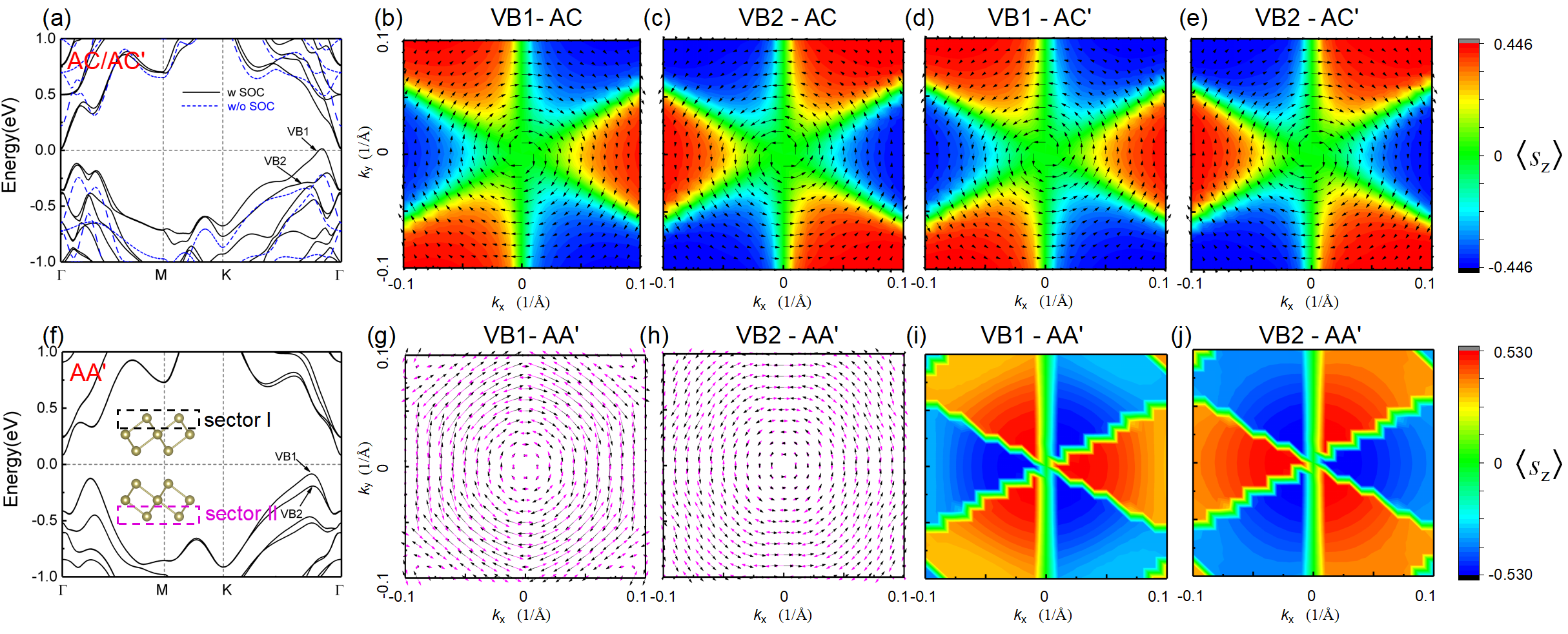}
    \caption{ (a) Band structure of AC/AC$^{\prime}$ phase with and without SOC effect. (b)-(e) The spin textures for highest valence bands (VB1 and VB2) for both AC and AC$^{\prime}$ phases.
    The black arrows represent the in-plane projection of the spin, while the background color indicates the out-of-plane projection of the spin.
    (f) Band structure of AA$^{\prime}$ phase considering SOC effect. The most top and bottom sublayers refer to sector I and II, respectively, as shown in the inset. (g)-(h) The layer-resolved in-plane spin polarization of AA$^{\prime}$ phase. The black arrow and magenta arrow represent for in-plane spin originated from sector-I and sector-II, respectively. (i)-(j)  The out-of-plane spin polarization of AA$^{\prime}$ phase.
    }
    \label{fig4}
\end{figure*}

The spin-orbit coupling (SOC) links the spin degree of freedom to the orbital motion of electrons in a solid and plays an important role in the emergence of new physical phenomena\cite{wolf2001spintronics}.
In particularly, for noncentrosymmetric materials, the SOC locks the electron's spin direction to its momentum resulting in non-trivial spin textures in the reciprocal space\cite{rashba1960properties}. In ferroelectric materials these spin textures are coupled to the ferroelectric polarization and thus can be controlled by its orientation and magnitude\cite{di2012electric}. This provides a promising platform to explore the coupling between spin, orbital, valley, and layer degrees of freedoms and opens a new direction for nonvolatile spintronic devices\cite{PhysRevB.105.035131}, such as a spin-field effect transistor and a valley spin valve.

The Fig.~\ref{fig4}(a) shows the the band structure of AC/AC$^{\prime}$ phase with SOC, which show a zero band gap nature under PBE level while the band gap is increased to 0.17 eV when considering HSE06 correction\cite{PhysRevB.93.224425}, indicating it is a narrow band gap semiconductor.
Moreover, the highest valence and the lowest conduction demonstrate substantial Rashba spin splitting (RSS) around $\Gamma$ point due to SOC\cite{PhysRevLett.102.056405,PhysRev.100.580}. For instance, the high valence band splits into two bands (VB1, VB2) with a maximum splitting amplitude of 0.22 eV along $\Gamma$K path.

In order to have a better understanding of the spin splitting nature and the relation between ferroelectric and spin polarization, the spin textures for VB1 and VB2 for AC/AC$^{\prime}$ phase are plotted in Figs.~\ref{fig4}(b)-(e).
Firstly, the in-plane spin components ($\langle \emph{s}_x\rangle$, $\langle \emph{s}_y\rangle$) exhibits opposite helical textures for VB1 and VB2, respectively, confirming the existence of Rashba-type splitting. Meanwhile the out of plane spin component ($\langle \emph{s}_z\rangle$) has three-fold symmetry in agreement with the three-fold rotation symmetry of the crystal. The direction of \emph{s}$_z$ is also opposite for VB1 and VB2.
Secondly, when flipping the electric polarization direction from up to down (from AC phase to AC$^{\prime}$ phase), the in-plane spin polarization is obviously switched, while the out-of-plane spin textures remain unchanged.
This phenomenon can be understood in the context of in-plane spin textures, taking into account the SOC effect. Electrons in crystal experiences an effective magnetic field  $ \textbf{B} = (\textbf{\emph{p}} \times \textbf{E} / 2mc^2 )$, where \textbf{\emph{p}}, \textbf{E}, m and c are the momentum, electric field, effective mass and velocity of light\cite{yuan2013zeeman}, respectively. Thus,
\textbf{B}$_{x/y}$ $\sim$ \textbf{\emph{p}}$_{y/x}$\textbf{E}$_{z}$ and \textbf{B}$_{z}$ $\sim$ \textbf{\emph{p}}$_{z}$\textbf{E}$_{x/y}$, this reveals that the out-of-plane electric polarization can alter the in-plane spin polarization while leaving the out-of-plane spin textures unaffected.

Moving on to the AA$^{\prime}$ stacking phase with $M_z$ symmetry, it also behaves as a semiconductor with an indirect band gap of 0.18 eV with SOC effect. In Fig.~\ref{fig4}(f), the highest valence band along the $\Gamma$-K path undergoes a spin splitting into two bands (VB1 and VB2) due to the SOC effect. Conventionally, due to constraint of $M_z$ symmetry the in-plane spin polarization of each band must be zero, with only net $\langle \emph{s}_z\rangle$ as shown in Figs.~\ref{fig4}(i)-(j). However, if one project a physical quantity e.g. $s$ into two different sectors which are linked by $M_z$ operation as shown in Figs.~\ref{fig4}(g)-(h),
it results in the relationship $\langle$\emph{s}$^{A}$$\rangle$$_{k}$= -$\langle$\emph{s}$^{B}$$\rangle$$_{k}$.
The phenomenon known as layer-dependent or hidden spin effect encompasses several aspects, including hidden spin polarization (HSP), hidden Berry curvature (HBC), and hidden valley polarization (HVP)\cite{PhysRevB.102.165143,PhysRevB.107.L081201,PhysRevB.107.245404,PhysRevB.105.205429}.
These effects may emerge in certain centrosymmetric non-magnetic materials with local symmetry breaking\cite{zhang2014hidden}.
Recent research studies have indicated that these hidden physical effects have been observed across a wide range of crystal symmetry groups\cite{PhysRevB.102.165143,PhysRevB.107.L081201}.
For the case of BL-Te, the top and bottom Te atoms are designated as sectors I and II, as illustrated in Fig.~\ref{fig4}(f).
It is observed that a clockwise spin texture is hosted by the top layer, while an anti-clockwise spin texture is hosted by the bottom layer for VB1. The presence of such layer-locked hidden spin polarization is likely to be probed in experiments using Scanning Tunneling Microscopy (STM)\cite{PhysRevLett.129.276601}. Interestingly, the orientations of spin textures projected onto sector A (or B) of VB1 and VB2 are both anticlockwise (clockwise), which differs from the traditional explicit Rashba spin texture where the spin direction of VB1 and VB2 is opposite.

\section{summary}
In summary, we present first-principles evidence of mono-elemental ferroelectricity in BL-Te with large SOC effect. The most stable stacking stacking patterns is obtained which can break the central symmetry, and thus induce OOP ferroelectric polarization with a magnitude of of 0.78 pC/m and small transition potential barrier of 29.2 meV. Due to large SOC strength and broken centrosymmetry, Rashha-like spin texture emerges with switchable spin polarization direction control by changing the electric polarization direction. In addition, the layer-dependent hidden spin polarization is also revealed in mirror-symmetric paraelectric phase. Our work provides theoretical guidance and candidate materials to explore interactions among polarization, layers, and spin degrees of freedom, offering theoretical guidance and candidate materials.

\begin{acknowledgements}
This work was supported by the Natural Science Foundation of China (Grant Nos. 22175150, 12204330, and U2002217) and the Key R$\&$D program of Yunnan Province (Grant No. 2018BA068), and the 14th graduate Research Innovation project of Yunnan University (KC-22221161). The numerical computations were performed at the Hefei advanced computing center.
\end{acknowledgements}

\end{document}